\documentstyle[12pt]{article}
\vspace{1.8cm} \oddsidemargin 0pt \evensidemargin 0pt  \topmargin
-1.5cm \footheight 26pt \footskip 10mm

\begin{document}
\baselineskip 20pt
\begin{center}
\baselineskip=24pt {\Large Dirac Equation at Finite Temperature}

\vspace{1cm} {Xiang-Yao Wu$^{a}$ \footnote{E-mail:wuxy2066@163.com
}, Bo-Jun Zhang$^{a}$, Xiao-Jing Liu$^{a}$, Nuo Ba$^{a}$\\ Yi-Heng
Wu$^{a}$, Si-Qi Zhang$^{a}$, Jing Wang$^{a}$ and Chun-Hong
Li$^{a}$} \vskip 10pt \noindent{\footnotesize a. \textit{Institute
of Physics, Jilin Normal University, Siping 136000, China}}

\end{center}
\date{}
\renewcommand{\thesection}{Sec. \Roman{section}} \topmargin 10pt
\renewcommand{\thesubsection}{ \arabic{subsection}} \topmargin 10pt
{\vskip 5mm
\begin {minipage}{140mm}
\centerline {\bf Abstract} \vskip 8pt
\par
\indent\\

\hspace{0.3in} In this paper, we propose finite temperature Dirac
equation, which can describe the quantum systems in an arbitrary
temperature for a relativistic particle of spin-1/2. When the
temperature $T=0$, it become Dirac equation. With the equation, we
can study the relativistic quantum systems in an arbitrary temperature.\\
\vskip 5pt
PACS numbers: 03.65.-w; 11.10.Wx\\

Keywords: Dirac equation; Finite temperature

\end {minipage}

\newpage
\section * {1. Introduction }

\hspace{0.3in} We know that the studies of physical systems at
finite temperature have led to many interesting properties such as
phase transitions, black body radiation etc. However, the study of
complicated quantum mechanical systems at finite temperature has
had a systematic development only in the past few decades. There
are now well developed and well understood formalisms to describe
finite temperature field theories, as they are called. In fact, as
we know now, there are three distinct, but equivalent formalisms
[1-3] to describe such theories and each has its advantages and
disadvantages. But, the important point to note is that we now
have a systematic method of calculating thermal averages in
quantum field theory. Quantum field theory at finite temperature
is the relativistic generalization of finite temperature
non-relativistic quantum statistical mechanics [4]. There it is
known that quantum effects are only important at low temperature.
More precisely the important parameter is the ratio of the thermal
wave-length and the length scale which characterizes the variation
of the potential. Only when this ratio is large are quantum
effects important. Increasing the temperature is at leading order
equivalent to decrease $\hbar$. Note that, from the point of view
of the path integral representation of quantum mechanics, the
transition from quantum to classical statistical mechanics appears
as a kind of dimensional reduction: in the classical limit path
integrals reduce to ordinary integrals.

With the finite temperature quantum field theories, we can now
study questions such as phase transitions involving symmetry
restoration in theories with spontaneously broken symmetry [5]. We
can study the evolution of the universe at early times which
clearly is a system at high temperature. More recently, even
questions such as the chiral symmetry breaking phase transition
[6] have drawn a lot of attention in view of the planned
experiments involving heavy ion collisions. This would help us
understand properties of the quark-gluon plasma better.

Quantum field theories at finite temperature are very challenging
also from the more theoretical point of view. There is a real-time
as well as an imaginary-time formalism, the first describing
dynamical and the second equilibrium properties [7]. Many
fundamental issues and problems are unsolved so far or require a
deeper understanding. Quantum field theories are subject to
enhanced complexities compared to zero temperature and zero
density. In this paper, we define the covariation microscopic
entropy for a relativistic particle of spin-1/2, and then obtain a
new finite temperature Dirac equation, which can study the
relativistic quantum systems in an arbitrary temperature, and can
further develop to become finite temperature quantum field theory.

\section * {2. The concept of microscopic and macroscopic entropy}

\hspace{0.3in} In thermodynamics, the infinitesimal entropy change
$dS$ of a system is defined by
\begin{equation}
dS=\frac{\delta Q}{T},
\end{equation}
where $\delta Q$ is a transfer of heat between the composite
system and an external reservoir at the temperature $T$. In a
finite change of state from $i$ to $j$, the entropy change is
$S_{j}-S_{i}$,
\begin{equation}
S_{j}-S_{i}=\int^{j}_{i}\frac{\delta Q}{T},
\end{equation}
The entropy of a system is a function of the thermodynamic
coordinates whose change is equal to the integral of $\frac{\delta
Q}{T}$ between the terminal states, integrated along any
reversible path connecting the two states.

By integrating (1) around a reversible cycle, so that the initial
and final entropies are the same. For a reversible cycle, we get
\begin{equation}
\oint\frac{\delta Q}{T}=0.
\end{equation}
In quantum statistics, the entropy $S$ is defined by
\begin{equation}
S=-k_{B}Tr(\rho \ln\rho),
\end{equation}
where $\rho=|\psi\rangle\langle\psi|$ is the density matrix,
$k_{B}$ is the Boltzmann constant.

In classical statistics, the entropy is defined by the Boltzmann
\begin{equation}
S=k_{B}\ln W=-k_{B}\ln\rho,
\end{equation}
where $W$ is the total number of the possible microscopic states,
and $\rho=\frac{1}{W}$ is probability of every state.

From (5), we know that the macroscopic entropy is from the state
distribution of a large number of particles. According to the
viewpoint of macroscopic entropy, as the single particle hasn't
state distribution, the single particle hasn't entropy.

In the viewpoint of quantum mechanics, a microscopic particle has
wave-particle duality, and the wave nature is described by wave
function. Obviously, the wave functions have the nature of
distribution. So, a single particle has microscopic entropy, and
we define its microscopic entropy $S$ as
\begin{equation}
S=-k_{w}\ln|\psi(\vec{r},t)|^{2}
\end{equation}
Where $k_{w}$ is a constant, which will be confirmed by
experiment.

For a relativistic particle of spin-1/2, we should consider the
covariation. In Eq. (6), we should replace the probability density
$|\psi|^{2}$ with Lorentz scalar $\bar{\psi}\psi$, and can obtain
the microscopic entropy $S$ for a relativistic particle of
spin-1/2, they are
\begin{equation}
S=-k_{w}\ln(\bar{\psi}\psi).
\end{equation}

\section * {3. The relation between a particle energy and temperature}
Based on the first law of thermodynamics, there is
\begin{equation}
dU=dW+dQ,
\end{equation}
where $dU$ is the infinitesimal change of system internal energy,
$dW$ is the infinitesimal work doing by surrounding, and $dQ$ is
the infinitesimal heat absorbed from surrounding. In this paper,
we should study the quantum thermodynamics property of microscopic
particles, i.e., finite temperature quantum theory of N-particle.
When $dW=0$, the (9) becomes
\begin{equation}
dU-dQ=0.
\end{equation}
For the point particle system of N-particle, the internal energy
$U$ is
\begin{equation}
U=\sum_{i=1}^{N}(T_{i}+V_{i})+\sum_{i<j}^{N}V_{ij},
\end{equation}
where $T_{i}$ and $V_{i}$ are the i-th particle's kinetic energy
and potential energy, and $V_{ij}$ is the interaction energy of
i-th and j-th particle, the internal energy infinitesimal change
is
\begin{equation}
dU=d(\sum_{i=1}^{N}(T_{i}+V_{i})+\sum_{i<j}^{N}V_{ij}),
\end{equation}
since
\begin{equation}
dQ=\sum_{i=1}^{N}dQ_{i}=\sum_{i=1}^{N}TdS_{i}=d(\sum_{i=1}^{N}TS_{i}),
\end{equation}
where $dQ_{i}$ is the i-th particle absorbed heat, $dS_{i}$ is the
i-th entropy change. $T$ is the system temperature. \\
By substituting (12), (13) into (10), we have
\begin{equation}
d(\sum_{i=1}^{N}(T_{i}+V_{i})+\sum_{i<j}^{N}V_{ij}-T\sum_{i=1}^{N}S_{i})=0.
\end{equation}
i.e.,
\begin{equation}
\sum_{i=1}^{N}(T_{i}+V_{i})+\sum_{i<j}^{N}V_{ij}-TS=E,
\end{equation}
where $S=\sum_{i=1}^{N}S_{i}$ is system total entropy, $E$ is a
constant, which can be defined as the total energy of system.

For the single particle, the (14) becomes
\begin{equation}
\frac{p^{2}}{2m}+V-TS=E,
\end{equation}
where $\frac{p^{2}}{2m}$, $V$ and $E$ are the single particle
kinetic energy, potential energy and total energy, $S$ is the
single particle microscopic entropy and $T$ is the single
particle's surrounding temperature. For a classical particle, its
entropy is zero, and its total energy $E=\frac{p^{2}}{2m}+V$,
i.e., its total energy is mechanical energy. For a microscopic
particle, due to the wave nature, it has microscopic entropy. The
microscopic particle total energy $E$ is the sum of its mechanical
energy $\frac{p^{2}}{2m}+V$ and $-TS$. We define $-TS$ as heat
potential energy.

For a relativistic free particle, the total squared energy of the
relativistic particle is
\begin{eqnarray}
E^{2}=m_{0}^{2}c^{4}+\vec{p}^{2}c^{2}.
\end{eqnarray}
For a relativistic unfree particle, including interaction
potential $V(r)$, its relativistic total energy and total squared
energy are
\begin{eqnarray}
E=mc^{2}+V(r),
\end{eqnarray}
and
\begin{eqnarray}
(E-V(r))^2=m_{0}^{2}c^{4}+\vec{p}^{2}c^{2},
\end{eqnarray}
when a particle is in temperature $T$, its non-relativistic total
energy is given by Eq. (15). Similarly, when a relativistic
particle is in temperature $T$, its relativistic total energy is
\begin{eqnarray}
E=mc^{2}+V(r)-TS,
\end{eqnarray}
and its total squared energy is
\begin{eqnarray}
(E-V(r)+TS)^2=m_{0}^{2}c^{4}+\vec{p}^{2}c^{2}.
\end{eqnarray}
\section * {4. Finite temperature Dirac Equation}
For the free particle of spin-1/2, we can obtain Dirac equation
from Eq. (16)
\begin{equation}
i\hbar\frac{\partial}{\partial t}\psi(\vec{r},t)=(-i\hbar c\vec
{\alpha}\cdot\nabla+\beta m_0c^2)\psi(\vec{r},t),
\end{equation}
where the $\alpha_{i}$ (i=1, 2, 3) and $\beta$ matrix are
\begin{equation}
\alpha_{i}=\left(%
\begin{array}{cc}
  0 & \sigma_{i} \\
\sigma_{i}& 0 \\
\end{array}%
\right),\hspace{0.3in} \beta=\left(%
\begin{array}{cc}
  I & 0 \\
  0 & I \\
\end{array}%
\right),
\end{equation}
and $\sigma_{i}$ are the usual $2\times2$ pauli matrix and $I$ the
$2\times2$ unit matrix, the wave function $\psi$ is
\begin{equation}
\psi=\left(%
\begin{array}{c}
  \psi_{1} \\
  \psi_{2} \\
  \psi_{3} \\
  \psi_{4} \\
\end{array}%
\right).
\end{equation}
Multiplied from the right by $\beta/c$, and with the definitions
\begin{equation}
\gamma^{0}=\beta,\hspace{0.3in}
\gamma^{i}=\beta\alpha_{i}\hspace{0.3in}(i=1, 2, 3).
\end{equation}
The Eq. (21) becomes
\begin{equation}
i\hbar(\gamma^{0}\frac{\partial}{\partial
x^{0}}+\gamma^{1}\frac{\partial}{\partial
x^{1}}+\gamma^{2}\frac{\partial}{\partial
x^{2}}+\gamma^{3}\frac{\partial}{\partial x^{3}})\psi-m_0c\psi=0,
\end{equation}
where
\begin{equation}
\gamma^{i}=\left(%
\begin{array}{cc}
  0 & \sigma_{i} \\
-\sigma_{i}& 0 \\
\end{array}%
\right),\hspace{0.3in} \gamma^{0}=\left(%
\begin{array}{cc}
  I & 0 \\
  0 & -I \\
\end{array}%
\right),
\end{equation}
The Eq. (26) can be written as
\begin{equation}
(i\hbar\gamma^{\mu}\partial_ \mu-m_0c)\psi=0,
\end{equation}
we set $\hbar=c=1$, the Eq. (28) is
\begin{equation}
(i\gamma^{\mu}\partial_ \mu-m_0)\psi=0.
\end{equation}
For the unfree particle of spin-1/2, we can obtain the new Dirac
equation from Eq. (10)
\begin{equation}
i\hbar\frac{\partial}{\partial t}\psi=(-i\hbar
c\vec{\alpha}\cdot\nabla+\beta m_0c^2+V(r)-TS)\psi,
\end{equation}
substituting Eq. (7) into (29), we have
\begin{equation}
i\hbar\frac{\partial}{\partial t}\psi=(-i\hbar
c\vec{\alpha}\cdot\nabla+\beta m_0c^2+V(r)+k_{w}T
\ln(\bar{\psi}\psi) \hspace{0.05in})\psi.
\end{equation}
The Eq. (30) can be simplied as
\begin{equation}
(i\gamma^{\mu}\partial_ \mu-m_0-\beta V-\beta k_{w}T
\ln\bar{\psi}\psi\hspace{0.05in})\psi=0,
\end{equation}
when $V=0$, the Eq. (31) becomes
\begin{equation}
(i\gamma^{\mu}\partial_ \mu-m_0-\beta k_{w}T
\ln(\bar{\psi}\psi)\hspace{0.05in})\psi=0.
\end{equation}
The Eq. (32) is the Dirac equations at finite temperature. When
$T=0$, it becomes Dirac equation, i.e., the Dirac equation is the
zero temperature relativistic wave equation of spin-1/2 particle.
\section * {4. Conclusion}
\hspace{0.3in}In this paper, we define microscopic entropy of a
single particle, and give the finite temperature Dirac equation.
With the equation, we can study the quantum systems in an
arbitrary temperature, such as it can be studied superconductivity
mechanism, finite temperature quantum field theory and so on.

\end{document}